\documentclass[10pt,letterpaper]{article}
\usepackage{opex3}

\begin{document}

\title{Field test of a practical secure communication network with
decoy-state quantum cryptography}

\author{Teng-Yun Chen$^{1}$, Hao Liang$^{1}$, Yang Liu$^{1}$, Wen-Qi Cai$^{1}
$, Lei Ju$^{1}$, Wei-Yue Liu$^{2}$, Jian Wang$^{1}$, Hao Yin$^{1}$, Kai Chen$%
^{1}$, Zeng-Bing Chen$^{1}$, Cheng-Zhi Peng$^{1}$, and Jian-Wei Pan$^{1}$}

\address{$^{1}$Hefei National Laboratory for Physical Sciences at Microscale and
Department of Modern Physics, University of Science and Technology of China,
Hefei, Anhui 230026, China\\
$^{2}$School of Information Science and Engineering, Ningbo University,
Ningbo, Zhejiang 315211, China}

\email{zbchen@ustc.edu.cn}

\begin{abstract}
We present a secure network communication system that operated with
decoy-state quantum cryptography in a real-world application scenario. The
full key exchange and application protocols were performed in real time
among three nodes, in which two adjacent nodes were connected by approximate
20 km of commercial telecom optical fiber. The generated quantum keys were
immediately employed and demonstrated for communication applications,
including unbreakable real-time voice telephone between any two of the three
communication nodes, or a broadcast from one node to the other two nodes by
using one-time pad encryption.
\end{abstract}



\section{Introduction}

Quantum cryptography can in principle offer the first provable unconditional
security between communication parties, which is guaranteed by fundamental
laws of quantum mechanics, rather than unproven computational assumptions.
The last two decades have witnessed dramatic advances in both theoretical
developments and successful experimental demonstration of quantum
cryptography systems, see, e.g., \cite%
{BB84,Gisin1997,Nishioka2002,Grosshans2005,GYS2004,Peng2005,Chen2006,Honjo2006,superconducting,Qiang08}
to name a few of them. Based on these attractive progresses, several
companies (such as IdQuantique, MagiQ and SmartQuantum) have commercially
developed quantum cryptography prototypes, which bring quantum cryptography
into practical applications by integrating with current encryption and
decryption techniques. In practice, however, the security of a specific
setup is not automatically ensured due to various imperfections. Most of
today's commercially available quantum cryptography systems rely on photon
sources from attenuated laser pulses, which forms a tremendous security
threaten for such systems. This is because weak coherent pulse sources
contain two or more photons per pulse with a non-vanishing probability,
leaving the systems susceptible a beam splitter attack from a formidable
eavesdropper. The photon number splitting attack is in fact the main
security threat of practical QKD schemes \cite{Huttner1995,BLMS2000}.
Rigorous security analysis on practical quantum key distribution (QKD)
system is proposed by Gottesman-Lo-L\"{u}tkenhaus-Preskill \cite{GLLP}. and
Inamori-L\"{u}tkenhaus-Mayers \cite{ILM}. However, the results are not
optimal, which can guarantee only a very limited key generation rates and
distances for a practical quantum cryptography system.

Recent revolutionary progress has been achieved by introducing the
idea of decoy state \cite{Hwang2003}, and by turning the idea into
systematical and rigorous theory and scheme in \cite{Wang2005} and
\cite{Lo-Ma-Chen2005}. By using decoy state within the common setup,
one can obtain much higher key generation rates and longer distances
(typically from less than 30 km, to more than 100 km), in the same
level compared with the case of using true single photon sources
\cite{Lo-Ma-Chen2005}. This leads to firstly successful experimental
demonstrations by Lo's group from Canada \cite{Lo-decoy} for 15 km,
and further for 60km \cite{Lo-decoy2}. Then implementations for more
than 100 km are almost simultaneously realized by research groups
from China \cite{Pan-decoy}, America \cite{America-decoy} and Europe
\cite{EU-decoy}. Also an implementation for 25.3 km is achieved by
Toshiba's group from UK \cite{Toshiba-decoy}.

So far most research groups all over the world have put forward QKD links,
however, operating in a point-to-point mode only, rather than networks with
multiple users. This has greatly restricted the domain of applicability of
quantum cryptography, which enjoys the extremely high security standard.
Subtle design and appropriate network topology are needed to be effectively
integrated into existing data networks to achieve a high key generation rate
and long distance for a secure communication network.

Phoenix \textit{et al.} \cite{Phoenix1995} proposed the idea of passive
quantum networks by using passive optical components, which can realize QKD
between one user to any other user in the network. Townsend \textit{et al.}
\cite{Townsend1994} demonstrated that QKD is feasible between any user to
any other one within a passive quantum network. However, photons are split
by couplers according to their ratio which nevertheless sacrifices greatly
the actual key generation rate. By using a network controller that actively
controls optical switches \cite{Townsend1997}, the first quantum
cryptography network, DARPA (The Defense Advanced Research Projects Agency )
quantum network, became operational since 2004 \cite{Elliott2002,Elliott2005}%
. One node contains an active 2-by-2 optical switch that can be used to
actively switch between two network topologies. This network currently links
the campuses of BBN Technologies, Harvard University and Boston University
(BU), with distances of approximately 10 km for both BBN-Harvard span and
BBN-BU span. In 2006 the NIST group also demonstrated a three-user active
quantum cryptography network with one transmitter using optical switches and
two receivers, each connected to transmitter by 1 km fiber links \cite%
{NIST-net}. Over 1 Mbps sifted-key rate was claimed to be generated in
either link. The European SECOQC (Secure Communication based on Quantum
Cryptography) quantum network \cite{SECOQC} has initiated since 2004 and
currently claims to have 4 nodes in Vienna city for a fiber ring network of
approximately $63$ km and one additional node which is $85$ km far from the
ring. It is based on the trusted relay paradigm \cite{SECOQC}. It mostly
focuses on an architecture allowing integration of heterogeneous QKD-link
devices. One node with decoy state device is also included in the tested
network. Recently Chen \textit{et al.} implemented a four-user quantum
cryptography network by taking star topology based on wavelength-division
multiplexing (WDM) \cite{Guo2007}. It was built in the commercial backbone
telecom fiber network in Beijing with the longest length of $42.6$ km for
fibers between two nodes.

The DARPA network \cite{Elliott2002,Elliott2005} realized the first quantum
cryptography network with 3 node, while the European SECOQC quantum network
\cite{SECOQC} gives the first implementation of integrated heterogeneous
QKD-link devices. At the mean time, the NIST network \cite{NIST-net} gives
very high sifted-key rate in a short distance network, while the quantum
network in Beijing \cite{Guo2007} gives longest length between two nodes.
These progresses are quite significant and represent big steps toward a
secure QKD network. However, there exist still big gaps from a practical
quantum cryptograph network. Without using decoy state, a prototype setup
cannot achieve secure distance of more than 30 km generally \cite%
{Lo-Ma-Chen2005,ILM} with the standard BB84 protocol. The implemented DARPA
network \cite{Elliott2005}, the NIST network \cite{NIST-net} and the network
realized in \cite{Guo2007} are all without using decoy states. Therefore,
these network, in fact, either are not secure, or cannot accomplish the
performance mentioned in their experiments. In addition, the distance for
secure network communication are quite short, namely, less than 10 km in the
case of DARPA network and only 1 km for NIST network.

\begin{figure}[ptb]
\begin{center}
\includegraphics[width=95mm]{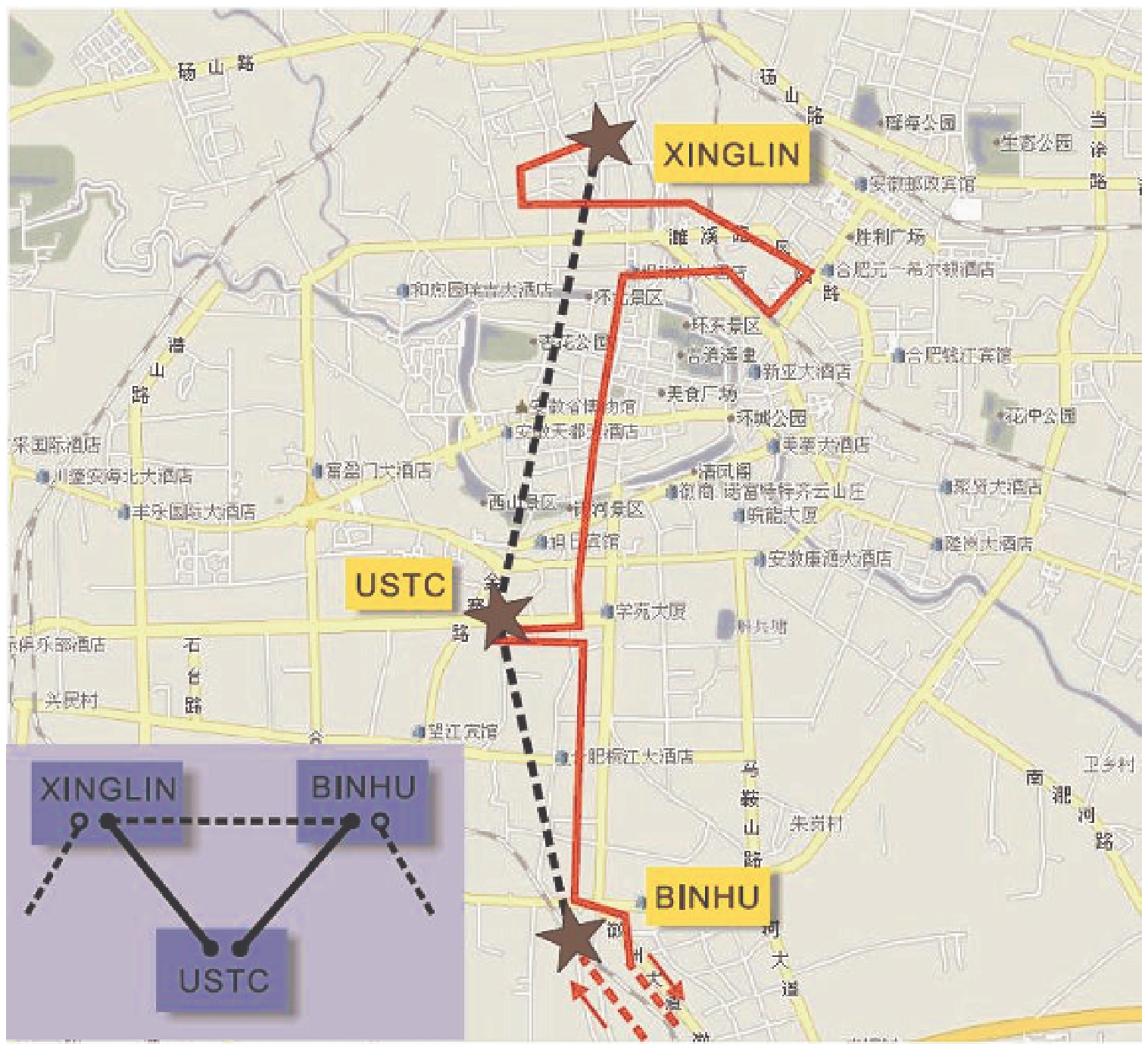}
\end{center}
\caption{Chained network architecture of our quantum cryptography network.
Two sets of decoy-state QKD systems are installed for Binhu-USTC link and
USTC-Xinglin link, respectively. The QKD systems have been updated in a
large degree to match seamless integration with real-time audio
communication by using one-time pad encryption, among the three nodes. The
red dashed line indicates the fiber running out of the map.}
\label{hfmap}
\end{figure}

In this article, we present a three-user network communication system based
on decoy-state quantum cryptography in a typical application scenario. In
the experiment, it is possible to create secure quantum keys on demand among
USTC (University of Science and Technology of China), Binhu, and Xinglin
that are located in Hefei city of China. As shown in Fig.~\ref{hfmap}, the
USTC node acts as a trusted relay and constitutes a chained QKD architecture
together with Binhu, and Xinglin. The telecom fiber strand is approximately
20 km for USTC-Binhu while it is also approximately 20 km for USTC-Xinglin.
The produced keys were directly handed over to an application that was used
to process real-time voice telephone between any two users of the three
nodes. We have developed secure communication network system including both
the QKD link modules and the audio application module based on quantum keys.
All of optical, electronical controlling, data acquisition and processing
system are integrated into one single box as a transmitter or a receiver.
Successful real-time secret audio communication has been performed between
any two users of the three nodes with the quantum keys through one-time pad
encryption \cite{one-time-pad}. An interphone has also been accomplished
when one implements a secure broadcast again by using one-time pad
encryption from one node to the other two nodes, or the other way around.

Compared with prior results, we provide a complete, compact, low cost 3-node
QKD network system in a real-life situation. Our motivation and results are
three-fold. Firstly we focus on a practical QKD network with decoy state.
The trusted-relay architecture we used is proved to be very practical and is
extensively used such as in DARPA network and the SECOQC network. It has
many advantages \cite{SECOQC} such as feasibility with today's technologies
(not relying on unavailability of quantum repeater), allowing for longer
distance compared with optical switch based network etc \cite{SECOQC}. At
the mean time, decoy state method can, in a large degree, increase the key
generation rate with guaranteed unconditional security. Secondly we have
focused on developing a complete system, with virtues of low cost and
compact, reliable and integrated components, rather than only an
experimental demonstration. This would help to bring a commercial QKD
network system closer. Thirdly, we focused on real-life application, such as
real time two-way audio communication and one-way broadcast, by utilizing
one-time pad encryption and decryption. The pseudo-random numbers are not
used in our system as they are normally used in a Gigahertz QKD system \cite%
{G-hertz}, due to lack of a random number generators in the Gigahertz level.
Rather we use the true random number generators in every place for the
system, and has achieved practical applications with unconditional security.
In addition, we use the InGaAs-type detectors with a small volume rather
than upconversion \cite{upconversion1,upconversion2} or superconducting
nanowire detectors \cite{superconducting}. The latter two detectors have the
advantage of high repetition rate for detection, but with a very large
volume. Moreover high background count rate is accompanied with the
upconversion detector, while superconducting detectors require cryogenic
cooling. The InGaAs-type detector thus provides an ideal choice for our
compact, low cost practical QKD network systems.

\section{Experimental setup}

Due to the currently lowest dispersion and attenuation for optical fiber at
the telecom wavelength, we implement our setup in a real-life situation by
using the running fiber network of China Netcom Group Corp Ltd. The laser
sources in our quantum cryptography system are produced from distributed
feedback (DFB) diodes with a center wavelength of $1550.12$ nm and pulse
duration of $1$ ns. By a random attenuation through a fiber intensity
modulator for the DFB laser, one thus creates the needed weak coherent
signal, decoy pulses and vacuum for this quantum cryptography setup. In our
system, there is a transmitter box in Binhu and a receiver box in Xinglin,
while in USTC there are both a transmitter box and a receiver box. Every box
has integrated full functions for control of QKD hardware, execution of QKD
protocol module and seamless interchange with our audio communication
application. This design can thus constitute two QKD links simultaneously
between Binhu and USTC, and between USTC and Xinglin. A repetition rate of 4
MHz is used for the laser source. This is because true random number
generators can only work at this level of rate for a commercially available
product. We adopt in this experiment phase encoding method for finishing QKD
tasks.

\begin{figure}[ptb]
\begin{center}
\includegraphics[width=100mm]{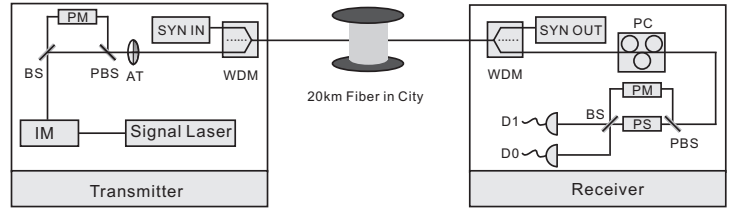}
\end{center}
\caption{Sketch of the experimental setup for one QKD-link. With a random
choice of measurement basis controlled by the phase modulator at the MZ
interferometer in Bob's side, Bob has 1/2 probability to have correct basis
choice. Both sides can then obtain sifted keys after comparison, which are
used for further error correction, privacy amplification according to decoy
state QKD mechanism. The classical communication channel is realized via a
standard TCP/IP connection in our setup. Here, IM (PM): intensity (phase)
modulator; BS: beam splitter; AT: controllable attenuator; SYN: synchronized
signal; PC: polarization controller; PS: phase shifter; D: single-photon
detector; PBS: polarizing beam splitter.}
\label{node-setup}
\end{figure}

For key generation, in transmitter's side the photon pulse is firstly sent
to an time division fiber Mach-Zehnder (MZ) interferometer with the long arm
through a phase modulator to generate the four primary signal states
necessary for implementing the BB84 \cite{BB84} protocol of a QKD system.
Here we use the polarizing beam splitters and beam splitters in both Alice's
and Bob's sides, such that photons from Alice's short arm are directed into
Bob's long arm and vice versa \cite{GYS2004}. This would avoid a 3dB loss
for useful photons in the normal case of Mach-Zehnder (MZ) interferometers
where only beam splitters are used. By using an attenuator through suitable
attenuation, one can control the photon number intensity to be $0.65/$pulse
for signal states, and $0.08/$pulse for decoy states for USTC-Xinglin link,
while they are $0.60/$pulse for signal states, and $0.20/$pulse for decoy
states for Binhu-USTC link. A synchronization laser pulse at the wavelength
of $1310$ nm was then combined with the signal and decoy states, through a
wavelength division multiplexing (WDM) apparatus, into the installed
single-mode telecom fiber for transmission. After passing through the $20$
km long dark fiber, at receiver's side the synchronization information
encoded in the fiber is firstly read out through another WDM apparatus.
Finally a clock signal synchronized with transmitter is formed, which will
further control correspondingly the measurement basis choice for the phase
modulator located in the unbalanced fiber MZ interferometer in receiver's
side. In addition, this synchronization clock signal will also act as the
gate control signal for the InGaAs-type detectors D0 and D1. The whole
synchronization electronics, detection logic and signal acquisition are all
integrated in a single board by using a field programmable gate array (FPGA)
and running at a sampling frequency of 4 MHz. The detectors are running in
gating mode while the gate width is set to be 2 ns to match our laser
source. For our detectors, the ``after pulse" will generally increase the
error rate of the raw key. The after pulse probability for the detectors
used in our setup will decrease to about 8/1000 if we set dead time be $%
20\mu s$. This is in the same level of duration for dead time if compared
with a recent experiment \cite{Gisin-Zbinden200809}, in which where a dead
time of $30\mu s$ is used. Thus we have set dead time for all detectors be $%
20\mu s$, and simultaneously match the detection events. The true dark
counts rate for the detectors themselves are all about $1.0\times 10^{-5}$%
/pulse. The measured value from vacuum decoy state for dark counts rate $Y_0$
is about $1.0\times 10^{-4}$/pulse due to finite extinction ratio for
intensity modulator, affect of the ``after pulse" for detectors, and the
intrinsic dark counts. The detection efficiency for all the detectors are
greater than $10\%$.

The telecom single mode fiber has an average attenuation of about $0.2$ dB
per kilometer resulting in a total attenuation of $4.5$ dB including the
connectors for Binhu and USTC span, and $5.6$ dB for USTC and Xinglin span.
To keep the identical and good coherent property for photons after
propagating along the long distance fiber, a voltage-driven fiber
polarization controller is used to dynamically adjust polarizations for the
transmitted states according to total detection rates. This active
compensation technique finally urges that photons from Alice's short arm are
directed into Bob's long arm and vice versa. To remove system's intrinsic
phase fluctuation in the MZ interferometers at both sides, we have used a
phase shifter to compensate the phase difference dynamically. This is
accomplished by implementing a feedback control to stabilize the phase.
Specifically we have inserted another pulsed laser (not shown in Fig.~\ref%
{node-setup}) with the same central wavelength of $1550.12$ nm as a
reference during the idle gap between two signal pulses, and making
continuous active control of the MZ interferometer arm lengths.

For satisfying the necessary requirements for decoy state QKD system, we
used true random number generators produced by IdQuantique (type:
Quantis-OEM, which has passed the NIST and Diehard randomness tests). These
random number generators are integrated in our controlling electronics: a)
to process random attenuation of laser source for producing signal and decoy
states; b) to load in the phase modulator in transmitter's side for
generating the needed four possible states for QKD system; and c) to load in
the phase modulator in receiver's side for forming the needed two possible
measurement basis for QKD system.

\section{Secure key generation and applications}

In this section, we present typical characters supplied by our experimental
network communication system. Besides the transmission loss in the fiber,
there are also other coupling and connection losses, in particular an
approximate 3.5 dB due to the inserting loss for the polarization
maintaining fiber in receiver's side. The BB84 protocol contributes an
additional 3 dB loss because there are only roughly one half of received
photons encoding correct information. It should be remarked that this loss
can be avoided if one uses an asymmetric basis choice for Alice and Bob \cite%
{Lo-Chau-Ardehali2005}. Our setting for the proportion of three transmitted
states is $6:1:1$ among the signal state, decoy state and vacuum state.

Before demonstrating our audio application among the three nodes, we
have run and measured the average specifications that our system can
achieve. Through a thirty-minute running, we have obtained
corresponding parameters for the two QKD links, and listed all the
related measurement and processing results in Table 1. The sifted
key rates are archived to be more than $10.5$ kbps for Binhu-USTC
link and more than $9.0$ kbps for USTC-Xinglin link. The quantum bit
error rate (QBER) is measured to be about $1.6\%$ for Binhu-USTC
link and about $1.4\%$ for USTC-Xinglin link. According to the
systematical theory of decoy state QKD
\cite{Wang2005,Lo-Ma-Chen2005,Wang2005PRA,Lo2005PRA}, we have
developed a data post-processing unit to finish both error
correction and privacy amplification in real time by considering
finite key length and statistical fluctuation. For the implemented
algorithms themselves, we are mainly based on the result from the
NIST group \cite{NISTECC}, by noting that there is no decoy state in
the NIST case. Currently we realize the algorithms using software,
while a FPGA implementation is more preferable in the future for
high speed QKD links. Consequently we can achieve a final secure key
rate of more than $1.5$ kbps for both links.

\begin{table}[tbh]
\caption{Measured specification for QKD network system}
\label{table1}\centering
\begin{tabular}[b]{cccccc}
\hline\hline
link & Communication wavelength & QBER & Sifted-key rate & Final key rate &
\\ \hline
Binhu-USTC & 1550.12nm & $\sim$ 1.6\% & $>$ 10.5 kbps & $>$ 1.6 kbps &  \\
USTC-Xinglin & 1550.12nm & $\sim$ 1.4\% & $>$ 9.0 kbps & $>$ 1.5 kbps &  \\
\hline\hline
\end{tabular}%
\end{table}

We obtain the following key generation rate by using the result of \cite%
{GLLP,Lo-Ma-Chen2005}
\begin{equation}
R\geq q\{-Q_\mu f(E_\mu) H_2(E_\mu)+Q_1[1-H_2(e_1)]\},  \label{R}
\end{equation}
where the subscript $\mu$ is the average photon number per signal in
signal states; $Q_\mu$ and $E_\mu$ are the measured gain and the
quantum bit error rate (QBER) for signal states, respectively; $q$
is an efficiency factor for the protocol. $Q_1$ and $e_1$ are the
unknown gain and the error rate of the true single photon state in
signal states. To achieve maximum possible key generation rate, the
decoy state method can estimate the lower bound of $Q_1$ denoting as
$Q_{1}^{L}$, and the upper bound of $e_1$ denoting as $e_{1}^U$.
Thus the decoy approaches could provide an unconditional security
\cite{Wang2005,Lo-Ma-Chen2005} for QKD systems. We follow here the
method developed in \cite{Lo2005PRA,Lo-decoy2} to estimate good
bounds for $Q_1$ and $e_1$, and using the stronger version for
maximizing the key generation rate formula developed in \cite%
{GLLP,Lo-Ma-Chen2005}. The $H_2(x)$ is the binary entropy function: $%
H_2(x)=-x\log_2(x)-(1-x)\log_2(1-x)$, while the factor $f(x)$ is for
considering an efficiency of the bi-directional error correction \cite%
{Brassard-Salvail1994}. For convenience, we denote $\nu$ the average photon
number per pulse for decoy state.

After experimentally measuring all the relevant parameters, we can input the
following bounds for calculating final key generation rate \cite%
{Lo2005PRA,Lo-decoy2}
\begin{eqnarray}
Q_1 \ge Q_1^L &=& \frac{\mu^2e^{-\mu}}{\mu\nu-\nu^2}(Q_\nu^L e^{\nu}-Q_\mu
e^\mu\frac{\nu^2}{\mu^2}-Y_0^U \frac{\mu^2-\nu^2}{\mu^2}), \\
e_1 \le e_1^U &=& \frac{E_\mu Q_\mu-Y_0^Le^{-\mu}/2}{Q_1^{L}},
\label{bounds}
\end{eqnarray}
in which
\begin{eqnarray*}
Q_\nu^L &=& Q_\nu(1-\frac{10}{\sqrt{N_\nu Q_\nu}}), \\
Y_0^L &=& Y_0(1-\frac{10}{\sqrt{N_0 Y_0}}), \\
Y_0^U &=& Y_0(1+\frac{10}{\sqrt{N_0 Y_0}}),  \label{pbounds}
\end{eqnarray*}
Here $N_\nu$, and $N_0$ are numbers of pulses used as decoy state and vacuum
state, respectively, while $Q_\nu$ is the measured gain for the decoy states.

All the relevant parameters are listed in Table \ref{table2} for a typical
running duration of 120s for USTC-Xinglin link in our experiment.
\begin{table}[h]
\caption{Measured and derived specification for decoy state system}
\label{table2}\centering
\begin{tabular}[b]{llllll}
\hline\hline
Para. & Value & Para. & Value &  &  \\ \hline
$Q_{\mu}$ & $6.36 \times10^{-3}$ & $E_{\mu}$ & $1.44 \times 10^{-2}$ &  &
\\
$Q_{\nu}$ & $8.61 \times10^{-4}$ & $E_{\nu}$ & $7.84 \times 10^{-2}$ &  &
\\
$Q_{1}^{L}$ & $2.72 \times10^{-3}$ & $e_{1}^U$ & $2.23 \times 10^{-2}$ &  &
\\
$R$ & $4.10 \times10^{-4}$ & $q$ & $0.356$ &  &  \\ \hline\hline
\end{tabular}%
\end{table}
From Table (\ref{table2}), we see a final key rate of around $4$M$*R=1.7$%
kbps is obtained for the typical running of our system. For achieving
unconditional security, we have estimated the bounds for $Q_{1}^{L}$ and $%
e_{1}^U$ by considering the statistical fluctuations for vacuum states,
gains for signal states and decoy states within 10 standard deviations. Thus
the final keys rates is valid for finite key length and promises a
confidence interval of about $1-1.5\times 10^{-23}$. We have performed
privacy amplification by utilizing the universal$_2$ functions that are
represented by Toeplitz matrices \cite{Krawczykhash}. This finally improves
both the efficiency and speed in a large degree for privacy amplification,
compared with the case that using purely random matrices. It should be
remarked that, to our knowledge, this is the first implementation for both
the error correction and privacy amplification, by considering statistical
fluctuation for decoy state quantum key distribution in a real-life
application. There is actually a tradeoff between key generation rate and
efficiency for privacy amplification. In our case we choose 120s
communication time for one time of executing privacy amplification for
corrected raw keys. One could certainly get faster realization for privacy
amplification for shorter communication time, then one is left with bigger
statistical fluctuation and thus less key generation rate.

We have accumulated a final key of about 120Mbits and performed the NIST
800-22 randomness test suite \cite{NIST800-22}. The sequence has passed all
the test for a significance level of 0.8\%, with the minimum pass rate for
each statistical test of 95\%. Also the Diehard statistical test suite \cite%
{Diehard} is performed. The reported $p$-value for the test are all between
0.009 and 0.989. Thus there is very high confidence of 98.9\% that our final
keys are truly random. With sufficient large data of keys, we hope to
perform more extensive random test for our system in the future.

Based on these results, we have developed telephony terminal equipments
through the normal analog commercial cable for telephony. The terminal has
an ability to make one-time pad encryption and decryption based on our QKD
links, to process common voice telephone. The audio compression ratio has
arrived $0.6$ kbps. In fact, our system can offer more than $1.2$ kbps,
which is two times the keys needed for a one-way communication. Thus our
system can offer directly two-way telephony communication in real time. We
have run the system for quite a few minutes, and always get clear audio
signal transmission with a good quality. After one hour's continuous
running, we still found no decrease of voice quality, which shows that our
setup provides a very stable and robust secure network communication system.
In fact, we have tested the whole system for half a month in USTC for a
telecom fiber of $20$ km. There is no any problem for the secure audio
communication system.

An interphone system is further developed in our experiment, which provides
a broadcast of ciphered information from one user to any other two users
with one-time pad encryption. Still using about a quantity of $0.6$ kbps
keys, we have successfully tested and finished broadcasts based on our
telephony system, from any one of the Binhu, USTC and Xinglin nodes to any
other two nodes. If we need feedbacks from the other two nodes, it is not
temporarily possible for all the nodes due to the limited key generation
rates. It is clear that it would need $1.2\times (N-1)$ kbps quantum keys to
process this task and to make all the two-way communications simultaneously
for $N$ nodes.

\section{Conclusion}

In summary, the experiment reported here demonstrates an operational network
communication system, which allows real-time voice telephone between any two
of the three communication users, or a broadcast from one user to the other
two users by using one-time pad encryption. The chained network topology
allows secret keys to be forwarded, in a hop-by-hop fashion, along QKD
links. Therefore unconditional authentication and encryption for information
transmission by using one-time pad will become possible. The middle node
acts as trusted relays and increases the key generation rate in a large
degree, compared with the case of direct connection between the nodes with
an exponential decreasing. Our setup can be easily expanded to many-node
network, and enjoys an advantage of slowly increasing for key's need. Near
future work would cover improving the key generation rates, by employing
high performance detectors and high-speed true random number generators etc.
We expect that it would be possible to finish two-way audio communications
in real time for QKD network with a few nodes. In the case that the key
rates is not enough for video conference with one-time pad encryption, we
expect to use classical symmetrical encryption algorithm such as AES
(Advanced Encryption Standard) with a high refreshing rate of keys, and
maintain a desired security level.

\section*{Acknowledgments}

We acknowledge the financial support from the CAS, the National Fundamental
Research Program of China under Grant No.2006CB921900, and the NNSFC. T.Y.C.
also gratefully acknowledges the support of China Postdoctoral Science
Foundation funded project and the K.C. Wong Education Foundation, Hong Kong
of China. The authors are grateful for very valuable discussions and
communications with Xiang-Bin Wang, Hoi-Kwong Lo, Bing Qi, Xiong-Feng Ma and
Yi Zhao.


\begin{thebibliography}{99}
\bibitem{BB84} C. H. Bennett and G. Brassard, \textquotedblleft Quantum
cryptography: public key distribution and coin tossing,\textquotedblright\
in \textit{Proceedings of the IEEE International Conferenceon Computers,
Systems and Signal Processing,} (Bangalore, India, 1984), pp. 175--179.

\bibitem{Gisin1997} A. Muller, T. Herzog, B. Huttner, W. Tittel, H. Zbinden,
and N. Gisin, \textquotedblleft `Plug and play' systems for quantum
cryptography,\textquotedblright\ Appl. Phys. Lett. \textbf{70,} 793--795
(1997).

\bibitem{Nishioka2002} T. Nishioka, H. Ishizuka, T. Hasegawa, and J. Abe,
\textquotedblleft `Circular type' quantum key
distribution,\textquotedblright\ Photon. Technol. Lett., IEEE \textbf{14,}
576--578 (2002).

\bibitem{Grosshans2005} F. Grosshans, G. V. Assche, J. Wenger, R. Brouri, N.
J. Cerf, and P. Grangier, \textquotedblleft Quantum key distribution using
Gaussian-modulated coherent states,\textquotedblright\ Nature \textbf{421,}
238--241 (2003).

\bibitem{GYS2004} C. Gobby, Z. L. Yuan, and A. J. Shields, \textquotedblleft
Quantum key distribution over 122 km of standard telecom
fiber,\textquotedblright\ Appl. Phys. Lett. \textbf{84,} 3762--3764 (2004).

\bibitem{Peng2005} C.-Z. Peng, T. Yang, X.-H. Bao, J. Zhang, X.-M. Jin,
F.-Y. Feng, B. Yang, J. Yang, J. Yin, Q. Zhang, N. Li, B.-L. Tian, and J.-W.
Pan, \textquotedblleft Experimental free-space distribution of entangled
photon pairs over 13 km: towards satellite-based global quantum
communication,\textquotedblright\ Phys. Rev. Lett. \textbf{94,} 150501
(2005).

\bibitem{Chen2006} T.-Y. Chen, J. Zhang, J.-C. Boileau, X.-M. Jin, B. Yang,
Q. Zhang, T. Yang, R. Laflamme, and J. W. Pan, \textquotedblleft
Experimental quantum communication without a shared reference
frame,\textquotedblright\ Phys. Rev. Lett. \textbf{96,} 150504 (2006).

\bibitem{Honjo2006} T. Honjo, K. Inoue, A. Sahara, E. Yamazaki, and H.
Takahashi, \textquotedblleft Quantum key distribution experiment through a
PLC matrix switch,\textquotedblright\ Opt. Commun. \textbf{263,} 120--123
(2006).

\bibitem{superconducting} H. Takesue, S.W. Nam, Q. Zhang, R.H. Hadfield, T.
Honjo, K. Tamaki, and Y. Yamamoto, ``Quantum key distribution over a 40-dB
channel loss using superconducting single-photon detectors", Nat. Photonics
\textbf{1,} 343--348 (2007).

\bibitem{Qiang08} Q. Zhang, H. Takesue, T. Honjo, K. Wen, T. Hirohata, M.
Suyama, Y. Takiguchi, H. Kamada, Y. Tokura, O. Tadanaga, Y. Nishida, M.
Asobe, and Y. Yamamoto, \textquotedblleft Megabits secure key rate quantum
key distribution,\textquotedblright\ eprint arxiv:quant-ph/0809.4018 (2008).

\bibitem{Huttner1995} B. Huttner, N. Imoto, N. Gisin, and T. Mor, ``Quantum
cryptography with coherent states", Phys. Rev. A \textbf{51,} 1863 (1995).

\bibitem{BLMS2000} G. Brassard, N. L\"utkenhaus, T. Mor and B.C. Sanders,
``Limitations on Practical Quantum Cryptography", Phys. Rev. Lett. \textbf{%
85,} 1330 (2000).

\bibitem{GLLP} D. Gottesman, H.-K. Lo, N. L\"utkenhaus, and J. Preskill,
``Security of quantum key distribution with imperfect devices", Quant. Inf.
Comput. \textbf{5,} 325--360 (2004).

\bibitem{ILM} H. Inamori, N. L\"{u}tkenhaus, D. Mayers, ``Unconditional
security of practical quantum key distribution", Eur. Phys. J. D \textbf{41,}
599-627 (2007).

\bibitem{Hwang2003} W. Y. Hwang, \textquotedblleft Quantum key distribution
with high loss: toward global secure
communication,\textquotedblright\ Phys. Rev. Lett. \textbf{91,}
057901 (2003).

\bibitem{Wang2005} X.-B. Wang, \textquotedblleft Beating the
photon-number-splitting attack in practical quantum
cryptography,\textquotedblright\ Phys. Rev. Lett. \textbf{94,}
230503 (2005).

\bibitem{Lo-Ma-Chen2005} H.-K. Lo, X.-F. Ma, and K. Chen, \textquotedblleft
Decoy state quantum key distribution,\textquotedblright\ Phys. Rev.
Lett. \textbf{94,} 230504 (2005); see also H.-K. Lo, Proceedings of
IEEE ISIT (International Symposium on Information Theory) 2004, p.
137 (IEEE Press, 2004).

\bibitem{Lo-decoy} Y. Zhao, B. Qi, X.-F. Ma, H.-K. Lo, and L. Qian,
\textquotedblleft Experimental quantum key distribution with decoy
states,\textquotedblright\ Phys. Rev. Lett. \textbf{96,} 070502 (2006).

\bibitem{Lo-decoy2} Y. Zhao, B. Qi, X.F. Ma, Hoi-Kwong Lo, L. Qian,
``Simulation and Implementation of Decoy State Quantum Key Distribution over
60km Telecom Fiber", Proceedings of IEEE International Symposium on
Information Theory 2006, pp. 2094-2098.

\bibitem{Pan-decoy} C.-Z. Peng, J. Zhang, D. Yang, W.-B. Gao, H.-X. Ma, H.
Yin, H.-P. Zeng, T. Yang, X.-B. Wang, and J.-W. Pan, \textquotedblleft
Experimental long-distance decoy-state quantum key distribution dased on
polarization encoding,\textquotedblright\ Phys. Rev. Lett. \textbf{98,}
010505 (2007).

\bibitem{America-decoy} D. Rosenberg, J. W. Harrington, P. R. Rice, P. A.
Hiskett, C. G. Peterson, R. J. Hughes, A. E. Lita, S. W. Nam, and J. E.
Nordholt, \textquotedblleft Long-distance decoy-state quantum key
distribution in optical fiber,\textquotedblright\ Phys. Rev. Lett. \textbf{%
98,} 010503 (2007).

\bibitem{EU-decoy} T. Schmitt-Manderbach, H. Weier, M. F\"{u}rst, R. Ursin,
F. Tiefenbacher, T. Scheidl, J. Perdigues, Z. Sodnik, C. Kurtsiefer, J. G.
Rarity, A. Zeilinger, and H. Weinfurter, \textquotedblleft Experimental
demonstration of free-space decoy-state quantum key distribution over 144
km,\textquotedblright\ Phys. Rev. Lett. \textbf{98,} 010504 (2007).

\bibitem{Toshiba-decoy} Z.L. Yuan, A.W. Sharpe, and A.J. Shields,
``Unconditionally secure one-way quantum key distribution using decoy
pulses", Appl. Phys. Lett. \textbf{90,} 011118 (2007).

\bibitem{Phoenix1995} S. J. D. Phoenix, S. M. Barnett, P. D. Townsend, and
K. J. Blow, \textquotedblleft Multi-user quantum cryptography on optical
networks,\textquotedblright\ J. Mod. Opt. \textbf{72,} 1155--1163 (1995).

\bibitem{Townsend1994} P. D. Townsend, S.J.D. Phonenix, K. J. Blow, and S.
M. Barnett, \textquotedblleft Quantum cryptography for multi-user passive
optical networks,\textquotedblright\ Electron. Lett. \textbf{30,} 1875--1877
(1994).

\bibitem{Townsend1997} P. D. Townsend, \textquotedblleft Quantum
cryptography on multi-user optical fibre networks,\textquotedblright\ Nature
\textbf{385,} 47--49 (1997).

\bibitem{Elliott2002} C. Elliott, \textquotedblleft Building the quantum
network,\textquotedblright\ New J. Phys. \textbf{4,} 46 (2002).

\bibitem{Elliott2005} C. Elliott, A. Colvin, D. Pearson, O. Pikalo, J.
Schlafer, and H. Yeh, \textquotedblleft Current status of the DARPA Quantum
Network,\textquotedblright\ in \textit{Quantum Information and Computation
III,} E. J. Donkor, A. R. Pirich, and H. E. Brandt, eds., Proc. SPIE \textbf{%
5815,} 138--149 (2005).

\bibitem{NIST-net} X. Tang, L.-J. Ma, A. Mink, A. Nakassis, H. Xu, B.
Hershman, J. Bienfang, D. Su, R. F. Boisvert, C. Clark, and C. Williams,
\textquotedblleft Demonstration of an active quantum key distribution
network,\textquotedblright\ in \textit{Quantum Communications and Quantum
Imaging IV,} R. E. Meyers, Y.-H. Shih, K. S. Deacon, eds., Proc. SPIE
\textbf{6305,} 630506 (2006).

\bibitem{SECOQC} SECOQC White Paper on Quantum Key Distribution and
Cryptography, http://www.secoqc.net/downloads/secoqc\_crypto\_wp.pdf,
accessed Feb. 2009.

\bibitem{Guo2007} W. Chen, Z.-F. Han, T. Zhang, H. Wen, Z.-Q. Yin, F.-X. Xu,
Q.-L. Wu, Y. Liu, Y. Zhang, X.-F. Mo, Y.-Z. Gui, G. Wei, and G.-C. Guo,
\textquotedblleft Field experimental `star type' metropolitan quantum key
distribution network,\textquotedblright\ eprint arxiv:quant-ph/0708.3546
(2007).

\bibitem{one-time-pad} G.S. Vernam, ``Cipher printing telegraph system for
secret wire and radio telegraph communications", J. Am. Inst. Electr. Eng.
\textbf{XLV,} 109--115 (1926).

\bibitem{G-hertz} Z. L. Yuan, A. R. Dixon, J. F. Dynes, A. W. Sharpe, and A.
J. Shields, \textquotedblleft Gigahertz quantum key distribution with InGaAs
avalanche photodiodes,\textquotedblright\ Appl. Phys. Lett. \textbf{92,}
201104 (2008).

\bibitem{upconversion1} T. Thew, S. Tanzilli, L. Krainer, S.C. Zeller, A.A.
Rochas, ``Low jitter up-conversion detectors for telecom wavelength GHz
QKD", I. Rech, S. Cova, H. Zbinden, and N. Gisin, New J. Phys. \textbf{8,}
32 (2006).

\bibitem{upconversion2} E. Diamanti, H. Takesue, C. Langrock, M.M. Fejer,
and Y. Yamamoto, ``100 km differential phase shift quantum key distribution
experiment with low jitter up-conversion detectors", Opt. Express \textbf{14,%
} 13073 (2006).

\bibitem{Gisin-Zbinden200809} D. Stucki, C. Barreiro, S. Fasel, J.-D.
Gautier, O. Gay, N. Gisin, R. Thew, Y. Thoma, P. Trinkler, F. Vannel, H.
Zbinden, ``High speed coherent one-way quantum key distribution prototype",
eprint arxiv:quant-ph/0809.5264 (2008).

\bibitem{Lo-Chau-Ardehali2005} H.-K. Lo, H.F. Chau, and M. Ardehali,
``Efficient Quantum Key Distribution Scheme and a Proof of Its Unconditional
Security", J. Cryptology \textbf{18,} 133 (2005)).

\bibitem{Wang2005PRA} X.-B. Wang, \textquotedblleft Decoy-state protocol for
quantum cryptography with four different intensities of coherent
light,\textquotedblright\ Phys. Rev. A \textbf{72,} 012322 (2005).

\bibitem{Lo2005PRA} X.-F. Ma, B. Qi, Y. Zhao, and H.-K. Lo,
\textquotedblleft Practical decoy state for quantum key
distribution,\textquotedblright\ Phys. Rev. A \textbf{72,} 012326 (2005).

\bibitem{NISTECC} A. Nakassis, J. C. Bienfang, and C. J. Williams,
\textquotedblleft Expeditious reconciliation for practical quantum key
distribution,\textquotedblright\ in \textit{Quantum Information and
Computation II,} A. R. Pirich, H. E. Brandt, eds., Proc. SPIE \textbf{5436,}
28--35 (2004).

\bibitem{Brassard-Salvail1994} G. Brassard, L. Salvail, \emph{Advances in
Cryptology EUROCRYPT '93}, Vol. 765 of \emph{Lecture Notes in Computer
Science}, (Springer, Berlin, 1994), pp. 410-423.

\bibitem{Krawczykhash} H. Krawczyk, ``LFSR-based Hashing and
Authentication", in \emph{Advances in Cryptology -- CRYPTO '94}, Y. G.
Desmedt, ed., Vol. 839 of \emph{Lecture Notes in Computer Science},
(Springer-Verlag London, 1994), pp. 129-139.

\bibitem{NIST800-22} http://csrc.nist.gov/groups/ST/toolkit/rng/index.html.

\bibitem{Diehard} G. Marsaglia, A. Zaman, W.W. Tsang, ``Toward a universal
random number generator", Stat. Prob. Lett., \textbf{9}, 35--39 (1990).
http://www.stat.fsu.edu/pub/diehard/, accessed Feb. 2009.
\end{thebibliography}
\end{document}